\documentclass[12pt]{iopart}
\usepackage{epsfig}
\usepackage{graphicx}
\usepackage{amssymb}

\newcommand{\bracket}[1]{\left\langle #1\right\rangle}

\newcommand{\be}{\begin{equation}}
\newcommand{\ee}{\end{equation}}
\newcommand{\bd}{\begin{displaymath}}
\newcommand{\ed}{\end{displaymath}}

\newcommand{\erf}{{\rm erf}}
\newcommand{\erfc}{{\rm erfc}}

\newcommand{\bbeta}{{\mbox{\boldmath $\eta$}}}

\newcommand{\bepsilon}{{\mbox{\boldmath $\epsilon$}}}

\newcommand{\unity}{{\mbox{\boldmath{1}\hspace{-1mm}\textbf{I}}}}

\begin{document}

\title[]{On the strategy frequency problem in batch Minority Games}

\author{A De Martino$^{1}$, I P\'erez Castillo$^{2,3}$, and D
Sherrington$^{3}$}

\address{$\,^{1}\,$CNR-INFM, Dipartimento di Fisica, Universit\`a di
  Roma ``La Sapienza'', p. le A. Moro 2, 00185 Roma, Italy}
\address{$\,^{2}\,$Dipartimento di Fisica, Universit\`a di Roma ``La
  Sapienza'', p. le A. Moro 2, 00185 Roma, Italy}
\address{$\,^{3}\,$Rudolf Peierls Centre for Theoretical Physics,
  University of Oxford, 1 Keble Road, Oxford, OX1 3NP, United Kingdom}
\ead{andrea.demartino@roma1.infn.it,isaac@thphys.ox.ac.uk,
  d.sherrington1@physics.ox.ac.uk}

\begin{abstract}
  Ergodic stationary states of Minority Games with $S$ strategies per
  agent can be characterised in terms of the asymptotic probabilities
  $\phi_a$ with which an agent uses $a$ of his strategies. We propose
  here a simple and general method to calculate these quantities in
  batch canonical and grand-canonical models. Known analytic theories
  are easily recovered as limiting cases and, as a further
  application, the strategy frequency problem for the batch
  grand-canonical Minority Game with $S=2$ is solved. The
  generalization of these ideas to multi-asset models is also
  presented. Though similarly based on response function techniques,
  our approach is alternative to the one recently employed by Shayeghi
  and Coolen for canonical batch Minority Games with arbitrary number
  of strategies.
\end{abstract}

\maketitle

\section{Introduction}

The mathematical theory of Minority Games (MGs) with $2$ strategies
per agent, particularly for what concerns their ergodic behaviour,
largely rests on the possibility of separating the contribution to
macroscopic quantities coming from ``frozen'' agents from that of
``fickle'' ones \cite{mgbook,coolen}.  Frozen agents are those who use
just one of their strategies asymptotically, whereas fickle agents
flip between their strategies even in the steady state.  That these
two groups have different impact on the physical properties of MGs is
clear if one thinks that frozen agents are insensitive to small
perturbations and thus they do not contribute to the susceptibility of
the system. More generally, when agents dispose of $S>2$ strategies
each, the relevant quantity to calculate is the probability with which
an agent uses $a$ of his strategies ($a\in\{0,1,\ldots,S\}$),
knowledge of which provides all interesting physical observables. On
the technical level, this is a rather complicated problem that has
been tackled only recently in \cite{shayeghi} for the canonical
$S$-strategy batch MG.  Here we propose an alternative method to
derive the desired statistics in generic canonical or grand-canonical
\cite{gcmg} settings with $S$ strategies per agent.  This approach has
the advantage of being simpler from a mathematical viewpoint and, as
we will show, easily exportable to other versions of the MG. As in
\cite{shayeghi}, we resort to path-integral techniques, allowing for a
description of the multi-agent dynamics in terms of the behavior of a
single, effective agent subject to a non-trivially correlated noise.
The central idea of the method we propose is to exchange the
integration over the effective noise for one over frequencies using a
simple invertible mapping from one set of variables to the other and
the transformation law of probability distributions. We show that
available theories are easily recovered in known cases and, as a
further application, solve the strategy frequency problem for the
grand-canonical MG with $S=2$.  Since a similar issue arises in the
context of multi-asset MGs \cite{ma}, we also discuss the
(straightforward though heavier from a notational viewpoint)
generalisation of this idea to models in which traders may invest in
$K\geq 2$ assets.

Since path integrals are by now a somewhat standard technique to deal
with MGs, we shall skip mathematical details and focus our analysis on
the resulting effective dynamics and specifically on the strategy
frequency problem. Moreover, we shall reduce the discussion of the
economic meaning of the model to the minimum. The interested reader
will find extensive accounts in \cite{mgbook,coolen,reviu}.

\section{Model definitions, TTI steady states and the strategy frequency problem}
\label{sect:model}

We consider a market for a single asset with $N$ agents, labeled by
$i\in\{1,\ldots,N\}\equiv\mathbb{Z}_N$. At each time step $\ell$,
agents receive an information pattern $\mu(\ell)\in\mathbb{Z}_P$
chosen randomly and independently with uniform probability and, based
on this, they formulate their bids (represented simply by a variable
encoding the agent's decision, e.g. to buy or sell the asset).  The
most interesting phenomenology is obtained when $P$ scales linearly
with $N$; their ratio, denoted as $\alpha=P/N$, is the model's main
control parameter. Every agent $i$ disposes of $S$ trading strategies
$\{a_{is}^{\mu}\}_{s\in\mathbb{Z}_S}$, each prescribing a binary
action $a^{\mu}_{is}\in\{-1,1\}$, drawn randomly and uniformly, and
independently for each strategy $s$ and pattern $\mu$. The performance
of every strategy is monitored by a score function $U_{is}(\ell)$
which is updated by
\begin{equation}\label{on-line}
U_{is}(\ell+1)-U_{is}(\ell)=- a_{is}^{\mu(\ell)}A(\ell)-\epsilon_{is}/\sqrt{N}
\end{equation}
Here, $\epsilon_{is}$ are real constants representing positive or
negative incentives for the agents to trade, with a factor $\sqrt{N}$
ensuring a non-trivial behavior in the limit $N\to\infty$. $A(\ell)$
is instead the (normalized) excess demand at time $\ell$,
\begin{equation}
A(\ell)=\frac{1}{\sqrt{N}}\sum_{i\in\mathbb{Z}_N} b_{i}(\ell)
\end{equation}
where $b_{i}(\ell)$ is the bid formulated by agent $i$ at time
$\ell$. If we denote by $s_{i}(\ell)$ the strategy chosen by $i$ at
time $\ell$, then the bid submitted by $i$ is given by
\begin{equation}
b_{i}(\ell)=\sum_{s\in\mathbb{Z}_S} n_{is}(\ell)a_{is}^{\mu(\ell)}
\delta_{s,s_{i}(\ell)}
\end{equation}
The terms $a_{is}^{\mu(\ell)}\delta_{s,s_i(\ell)}$ impose that the
agent performs the action dictated by his selected strategy. The term
$n_{is}(\ell)\equiv F[U_{is}(\ell)]$, with
$F:\mathbb{R}\to\mathcal{I}$, denotes a filter linked to the score of
the selected strategies. We focus our attention on two cases:
\begin{itemize}
\item Taking $F$ to be the Heaviside function, one has
$\mathcal{I}=\{0,1\}$ so that the filter consists in either submitting
($n_{is_i(\ell)}=1$ for $U_{i s_i(\ell)}>0$) or not submitting
($n_{is_i(\ell)}=0$ for $U_{i s_i(\ell)}<0$) the bid. This version of
the game is usually called grand-canonical MG \cite{gcmg}.
\item If $F\equiv 1$, the filter is absent and agents are forced to
play no matter how bad their scores perform. This corresponds to the
standard canonical MG.
\end{itemize}
It remains to describe how $s_{i}(\ell)$ is chosen. We assume
generically that at each time step agent $i$ employs a rule described
by a function $g_i$, namely
\begin{eqnarray}
s_{i}(\ell)= g_i\left[\{U_{is}(\ell)\}_{s\in\mathbb{Z}_S}\right]
\end{eqnarray}
For example, the standard MG with $S=2$ corresponds to $s_i(\ell)={\rm
  arg~}\max_{s\in\mathbb{Z}_S} U_{is}(\ell)$. (This generalises easily
to the case of traders with decision noise \cite{noiz}.) At this
stage, we assume that the $g_i$'s are chosen randomly and
independently across agents (from some distribution) and introduce the
density of the mappings $\{g_i\}_{i\in\mathbb{Z}_N}$ as
\begin{equation}\label{densa}
W[g]=\frac{1}{N}\sum_{i\in\mathbb{Z}_N}\delta_{(F)}(g- g_i)
\end{equation}
with $\delta_{(F)}(\cdots)$ a functional Dirac delta. A similar random
choice is made for incentives (albeit in general with a different and
uncorrelated distribution) and we define their density as
\begin{equation}
w(\bepsilon)=\frac{1}{N}\sum_{i\in\mathbb{Z}_N}
\prod_{s\in\mathbb{Z}_S}\delta\left (\epsilon_{s}-\epsilon_{is}\right)
\end{equation}
with $\bepsilon=\{\epsilon_{s}\}_{s\in\mathbb{Z}_S}$.

We will work out the `batch' version of the model, which is obtained
by averaging (\ref{on-line}) over information patterns
\cite{batch}. After a time re-scaling (we denote the re-scaled time as
$t$), one obtains the `batch' dynamics
\begin{eqnarray}
\fl U_{is}(t+1)-U_{is}(t)=\theta_{i}(t)-\alpha
\epsilon_{is}-\frac{1}{\sqrt{N}}\sum_{\mu\in\mathbb{Z}_P}
a_{is}^{\mu}\frac{1}{\sqrt{N}}\sum_{s\in\mathbb{Z}_S}\sum_{j\in\mathbb{Z}_N}
n_{js}(t)a_{js}^{\mu}\delta_{s,s_j(t)}\label{batch}
\end{eqnarray}
where $\theta_{i}(t)$ is a (small) external perturbation added for
later use. In dynamical studies, one is interested in the average bid
autocorrelation function
\begin{equation}
C(t,t')=\frac{1}{N}\sum_{i\in\mathbb{Z}_N}
\left[\bracket{b_{i}(t)b_{i}(t')}\right]_{dis}
\end{equation}
and in the average response function
\begin{equation}
G(t,t')=\frac{1}{N}\sum_{i\in\mathbb{Z}_N}
\left[\frac{\partial\bracket{b_{i}(t)}}{
\partial\theta_{i}(t')}\right]_{dis}
\end{equation}
where $\bracket{\cdots}$ and $[\cdots]_{dis}$ denote, respectively,
averages over paths and disorder.  Assuming that
$\theta_i(t)=\theta(t)$ for all $i$, in the limit $N\to\infty$ the
multi-agent dynamics (\ref{batch}) can be described in terms of a
self-consistent stochastic process for a single, effective agent
endowed with $S$ strategies, characterized by score functions
$U_s(t)$, ``spin'' variable
$s(t)=g\left[\{U_{s}(t)\}_{s\in\mathbb{Z}_S}\right]$ and filter
$n_s(t)=F[U_s(t)]$. This process can be derived by introducing a
generating function of the original dynamics and averaging over
disorder \cite{dedo}. Details of the calculation follow closely those
of similar models reported in the literature (see e.g. \cite{coolen}).
The effective dynamics ultimately reads
\begin{eqnarray}
\fl U_{s}(t+1)&=&U_{s}(t)+ \theta(t)-\alpha\epsilon_{s}
-\alpha\sum_{t'\leq t} [\unity+G]^{-1}(t,t')n_{s}(t')
\delta_{s(t'),s}+\eta_{s}(t)\,,
\end{eqnarray}
where $\eta_{s}(t)$ is a coloured Gaussian noise with first moments
given by
\begin{eqnarray}
\bracket{\eta_{s}(t)}_\star=0\\
\bracket{\eta_{s}(t)\eta_{s'}(t')}_\star
=\delta_{s,s'}\alpha[(\unity+G)^{-1} C (\unity+G^\dag)^{-1} ](t,t')
\end{eqnarray}
and where
\begin{eqnarray}
 C(t,t')=\sum_{s\in\mathbb{Z}_S} \int d\bepsilon\, w(\bepsilon)\int dg
\,W[g]\, \bracket{n_{s}(t)n_{s}(t')\delta_{s,s(t)} \delta_{s,s(t')}
}_\star\label{c-gen}\\ G(t,t')=\sum_{s\in\mathbb{Z}_S}\int d\bepsilon\,
w(\bepsilon) \int dg \,W[g]\,
\frac{\delta\bracket{n_{s}(t)\delta_{s,s(t)} }_\star}{\delta
\theta(t')}\label{chi-gen}
\end{eqnarray}
are the correlation and response functions, respectively.

We focus henceforth on ergodic steady-state properties, and more
precisely on time-translation invariant (TTI) solutions of
(\ref{c-gen}) and (\ref{chi-gen}). To do so we require that (a)
two-time quantities are Toeplitz-type matrices, {\it i.e.}
$C(t,t')=C(t-t')$, $G(t,t')=G(t-t')$, and that (b) there is no
anomalous integrated response, {\it i.e.}
$\chi:=\lim_{\tau\to\infty}\sum_{t\leq \tau} G(t)<\infty$. We denote
time-averages as
\begin{equation}
\overline{x}=\lim_{\tau\to\infty}\frac{1}{\tau}\sum_{t=1}^{\tau}x(t)
\end{equation}
Rewriting the scores as $U_{s}(t)=t u_{s}(t)$ and averaging over time
we obtain
\begin{equation}\label{ss}
u_{s}=\overline{\theta}+\overline{\eta}_{s}-
\alpha\epsilon_{s}-m\sum_{n\in\mathcal{I}} n f_{ns}
\end{equation}
where we have defined $m\equiv \frac{\alpha}{1+\chi}$,
$u_{s}=\lim_{\tau\to\infty} u_{s}(\tau)$ and
\begin{equation}
f_{ns}=\lim_{\tau\to\infty} \frac{1}{\tau}\sum_{t=0}^{\tau-1}
\delta_{n,n_{s}(t)}\delta_{s(t),s}
\end{equation}
In what follows, we set $\overline{\theta}=0$ (the response function
can be equally evaluated by a derivative with respect to the effective
noise $\overline{\eta}_s$). Note that (\ref{ss}) describes an ensemble
of processes, since in the stationary limit the noise variables
$\{\overline{\eta}_{s}\}_{s\in\mathbb{Z}_S}$ are Gaussian distributed,
\textit{viz.}
\begin{equation}
P(\overline{\bbeta})=\prod_{s\in\mathbb{Z}_S}
\frac{1}{\sqrt{2\pi\varsigma^2}}
\exp\left[-\frac{\overline{\eta}^2_{s}}{2\varsigma^2}\right] \,,\quad
\varsigma^2= \frac{\alpha c}{(1+\chi)^2}
\end{equation}
where the persistent autocorrelation $c=\lim_{\tau\to\infty}(1/\tau)
\sum_{t\leq\tau}C(t)$ and susceptibility $\chi$ can be computed
through
\begin{eqnarray}
c=\sum_{s\in\mathbb{Z}_S} \sum_{n,n'\in\mathcal{I}}n\,n'\int
d\bepsilon\, w(\bepsilon) \int dg\,W[g]\,
\bracket{f_{ns}f_{n's}}_{\star}\\ \chi=\frac{1}{\varsigma^2}
\sum_{s\in\mathbb{Z}_S}\sum_{n\in\mathcal{I}} n\int d\bepsilon\,
w(\bepsilon) \int dg\,W[g]\, \bracket{\overline{\eta}_{s}
f_{ns}}_{\star}
\end{eqnarray}
The coefficients $\{f_{ns}\}_{n\in\mathcal{I},s\in\mathbb{Z}_S}$ have
the meaning of frequencies. Indeed, $f_{ns}$ is the frequency of use
of strategy $s$ when the filter takes the value $n$. Clearly,
\begin{equation}
\sum_{n\in\mathcal{I}}\sum_{s\in\mathbb{Z}_S} f_{ns}=1
\end{equation}

Equation (\ref{ss}) is the staring point of our analysis.The problem
consists specifically in calculating the statistics of the frequency
variables.  For the sake of clarity, we shall now work out the
mathematical details of the strategy frequency problem in the case
recently addressed in the literature, namely that of the canonical MG
($F\equiv 1$) with $S$ strategies \cite{shayeghi}. Following sections
will address more complicated versions of the model.

\section{Canonical batch Minority Game with $S$ strategies}
\label{sect:MG_S_strategies}

Recalling that for canonical models $n=1$, in this section we simplify
the notation and write $f_s$ in place of $f_{ns}$.  Furthermore, in
order to make direct contact with the case discussed in
\cite{shayeghi}, we assume that $\epsilon_{s}=0$ for each
$s\in\mathbb{Z}_S$ and that the density $W[g]$ is a
$\delta$-distribution with
\begin{equation}
s(t)=g[\{U_s(t)\}]={\rm arg\, max}_{s\in\mathbb{Z}_S} U_{s}(t)
\label{eq:rule_case1}
\end{equation}
The stationary state equations now greatly simplify: for each $s$ we
have
\begin{eqnarray}
u_{s}&=&\overline{\eta}_{s}-m f_{s}\,,\quad m\equiv
\frac{\alpha}{1+\chi}\,,\quad \sum_{s\in\mathbb{Z}_S} f_{s}=1
\label{eq:SS_case1}
\end{eqnarray}
where $f_s$ is the frequency of use of strategy $s$. The statistics of
the frequencies can be evaluated as follows. Consider the case in
which the effective agent uses a subset of strategies
$\mathcal{A}\subseteq \mathbb{Z}_S$ ($\mathcal{A}\neq\emptyset$). Due
to the rule \eref{eq:rule_case1} this automatically implies that
\begin{eqnarray}
  u_s&=&u\,,\quad{\rm for~} s\in\mathcal{A}\,,\\ u_s&<&u\,,\quad{\rm
    for~} s\not\in\mathcal{A}
\end{eqnarray}
with $u$ a generic value of the score velocity. In turn, one has that
$\sum_{s\in\mathcal{A}}f_s=1$, the rest of the frequencies being
identically zero. Let us split the Gaussian variables in two groups:
\begin{equation}
\overline{\eta}_s=
\cases{x_s&for $s\in\mathcal{A}$\\
y_s&for $s\not\in\mathcal{A}$}
\end{equation}
We have
\begin{eqnarray}
&x_{s}\equiv x_s\left(u,\{f_s\}_{s\in\mathcal{A}}\right)\,, &\quad{\rm
for~}s\in\mathcal{A}\label{eq:eq_u_pos_case1}\\ &y_{s}<u\,,&\quad{\rm
for~} s\not \in\mathcal{A}
\end{eqnarray}
where $x_s\left(u,\{f_s\}_{s\in\mathcal{A}}\right)\equiv u+m f_s$ The
family of equations \eref{eq:eq_u_pos_case1} defines an invertible
mapping $\{x_s\}_{s\in\mathcal{A}}\to (u,\{f_s\}_{s\in\mathcal{A}})$
whose Jacobian is given by
\begin{equation}
\left|\frac{\partial \{x_s\}_{s\in\mathcal{A}}}{ \partial
(u,\{f_s\}_{s\in\mathcal{A}})}\right|=|\mathcal{A}|m^{|\mathcal{A}|-1}
\end{equation}
where $|\mathcal{A}|$ is the cardinality of $\mathcal{A}$.  We now
have all the information required to compute the frequency
distribution in this case. By simply invoking the transformation law
of probability distribution for the $\mathbf{x}$-variables,
\textit{i.e.}
\begin{equation}
P(\textbf{x}) d\textbf{x}=\varrho(u,\textbf{f})du\,
d\textbf{f}
\end{equation}
from whence
\begin{equation}
\varrho(u,\textbf{f})\equiv
P[\textbf{x}(u,\textbf{f})]\left|\frac{\partial \textbf{x}}{\partial
(u,\textbf{f})}\right|\,,\quad \textbf{f}=\{f_s\}_{s\in\mathbb{Z}_S}
\end{equation}
and the restriction over the distribution of the
$\textbf{y}$-variables, we have that the contribution to the frequency
distribution of the subset $\mathcal{A}$ of strategies with score $u$,
denoted $\varrho_{\mathcal{A}}(u,\textbf{f})$, reads
\begin{eqnarray}
\fl \varrho_{\mathcal{A}}(u,\textbf{f})=
|\mathcal{A}|m^{|\mathcal{A}|-1}
\delta\left(\sum_{s\in\mathcal{A}}f_s-1\right)
\left[\prod_{s\not\in\mathcal{A}}\delta_{f_s,0}\right]
P[\textbf{x}(u,\{f_s\}_{s\in\mathcal{A}})]
\bracket{\prod_{s\not\in \mathcal{A}}
\Theta\left(u-y_{s}\right)}_{\textbf{y}}
\end{eqnarray}
where we have used the fact that the noise distribution factorises,
\textit{i.e.} $P(\bbeta)=P(\textbf{x})P(\textbf{y})$ and emphasised
through the Dirac $\delta$-distributions the constraints over the
frequencies\footnote{We consider Dirac delta contributions coming from
the boundary of the integration region to be unity.}.
$\bracket{\cdots}_{\textbf{y}}$ denotes instead average over the
statistics of the $\textbf{y}$-variables.

Now the whole frequency distribution is simply given by the sum over
all possible partitions of $\mathbb{Z}_S$ (empty set not
included). Thus the average over the initial set of Gaussian variables
is converted to average over the frequency distribution:
\begin{equation}
\bracket{(\cdots)}_\star=
\sum_{\mathcal{A}\subseteq\mathbb{Z}_S|\mathcal{A}\neq \emptyset}\int
du\, d\textbf{f}\,\varrho_{\mathcal{A}}(u,\textbf{f})\,(\cdots)
\end{equation}

A further simplification is allowed here if one restricts the
attention to subsets with $|\mathcal{A}|=a$ by considering the
frequency distribution of $a$ strategies. By standard application of
combinatorics, one has
\begin{eqnarray}
\fl \varrho_{a}(u,\textbf{f})= \frac{S!}{(a-1)!(S-a)!}
\,m^{a-1} \delta\left(\sum_{s\in\mathbb{Z}_a}f_s-1\right) \left[
\prod_{s\not\in\mathbb{Z}_a}\delta_{f_s,0}\right] P[\textbf{x}(u,\{
f_s\}_{s\in\mathbb{Z}_a})]\nonumber\\
\times\bracket{\prod_{s \not\in
\mathbb{Z}_a}\Theta\left(u-y_{s}\right)}_{\textbf{y}}
\end{eqnarray}
and, in turn,
\begin{eqnarray}
\bracket{(\cdots)}_\star=\sum_{a\in\mathbb{Z}_S}\bracket{
(\cdots)_{a}}_{\star}&=&\sum_{a\in\mathbb{Z}_S} \int du\int
d\textbf{f}\,\varrho_{a}(u,\textbf{f}) (\cdots)_1
\end{eqnarray}
Now if we denote by $\phi_a$ the fraction of agents using $a$
strategies, then
\begin{equation}
\phi_a=\int du\, d\textbf{f}\, \varrho_{a}(u,\textbf{f})
\end{equation}
It easy to see that for $\phi_1$ and $\phi_2$ we obtain
\begin{eqnarray}
\fl \phi_1&=&S\int \frac{du}{\sqrt{2\pi\varsigma^2}}
e^{-\frac{(u+m)^2}{2\varsigma^2}}\left[\frac{1}{2}+\frac{1}{2}{\rm
Erf}\left(\frac{u}{\sqrt{2\varsigma^2}}\right)\right]^{S-1}\\
\fl\phi_2&=&S(S-1) m\int_0^1 \frac{df}{\sqrt{2\pi
\varsigma^2}}\int \frac{du}{\sqrt{2\pi\varsigma^2}} e^{-\frac{(u+m
f)^2}{2\varsigma^2}}e^{-\frac{[u+m(1-f)]^2}{2\varsigma^2}}
\left[\frac{1}{2}+\frac{1}{2}{\rm
Erf}\left(\frac{u}{\sqrt{2\varsigma^2}}\right)\right]^{S-2}
\end{eqnarray}
which, after some straightforward manipulations, is identified with
the corresponding formulas of \cite{shayeghi}. 

\section{Grand-canonical MG with one asset and $S$ strategies}
\label{sect:GCMG}

We now turn our attention to the grand-canonical version of the MG
with $S>1$ strategies strategies per agent. This is obtained by
taking, in addition to the rules used in the previous section,
$n_{is}(t)=F[\{U_{is}(t)\}]=\Theta[U_{is}(t)]$ instead of $F\equiv 1$.
Now the stationary state equations read, for each $s\in\mathbb{Z}_S$,
\begin{eqnarray}
u_{s}&=&\overline{\eta}_{s}-\alpha\epsilon_{s}-m
f_{s}\,,\quad m\equiv \frac{\alpha}{1+\chi}\,,\quad
\varphi+\sum_{s\in\mathbb{Z}_S}f_{s}=1
\label{eq:SS}
\end{eqnarray}
where we set $f_{1s}=f_s$ and denoted by $\varphi$ the probability
that the agent is inactive, that is the probability that $n=0$. In
this case the value of the frequencies for $n=0$ do not enter in the
relevant equations which determine the quantities of interest of the
model. We proceed to calculate the statistics of the frequencies
$\{f_s\}_{s\in\mathbb{Z}_S}$ and to relate all quantities to such
statistics. As before, let $\mathcal{A}\subseteq\mathbb{Z}_S$ be a
subset of strategies being used, so that
\begin{eqnarray}
u=u_s\,,\quad\forall s\in\mathcal{A}\\
u>u_s\,,\quad \forall s\not\in\mathcal{A}
\end{eqnarray}
Now we must distinguish three cases: if $u>0$, the agent is always
active, that is $\varphi=0$; if instead $u=0$, the agent is sometimes
inactive, that is $\varphi>0$; finally if $u<0$ then $\varphi=1$ and
the agent never invests.
\begin{enumerate}
\item \underline{Case $u>0$}. Here the analysis follows closely the
one performed for the canonical $S$-strategy MG. The agent is in the
market and $f_s\in[0,1]$ represents the frequency of the strategy $s$
being used. This implies that $f_s\neq 0$, $\forall s\in\mathcal{A}$
and $f_s=0$ $\forall s\not\in\mathcal{A}$ with
$\sum_{s\in\mathcal{A}}f_s=1$. We then split the stationary equations
\eref{eq:SS} into two parts and write
\begin{eqnarray}
&x_{s}=x^{+}_s\left(u,\{f_s\}_{s\in\mathcal{A}}\right)\,,&\quad\forall
s\in\mathcal{A}\label{eq:eq_u_pos}\\
&y_{s}<u+\alpha\epsilon_{s}\,,&\quad\forall s\not \in\mathcal{A}
\end{eqnarray}
where we have defined the functions
\begin{equation}
x^{+}_s\left(u,\{f_s\}_{s\in\mathcal{A}}\right)\equiv u+m
f_s+\alpha\epsilon_{s}
\end{equation}
and, as before, denoted as $\{x_s\}$ the Gaussian variables in the
subset $\mathcal{A}$ and as $\{y_s\}$ those not belonging to this
subset. The set of equations \eref{eq:eq_u_pos} defines an invertible
mapping $\{x_s\}_{s\in\mathcal{A}}\to (u,\{f_s\}_{s\in\mathcal{A}})$
whose Jacobian reads
\begin{equation}
\left|\frac{\partial \{x^{+}_s\}_{s\in\mathcal{A}}}{\partial
(u,\{f_s\}_{s\in\mathcal{A}})}\right|= |\mathcal{A}|
m^{|\mathcal{A}|-1}\,,
\end{equation}
with $|\mathcal{A}|$ the cardinality of the subset
$\mathcal{A}$. Proceeding as in the previous section, that is using
the transformation law of probability distributions, we find that the
contribution to the the frequency distribution of the subset
$\mathcal{A}$ of strategies, denoted
$\varrho_{\mathcal{A}}(u>0,\textbf{f})$, reads
\begin{eqnarray}
\fl \varrho_{\mathcal{A}}(u>0,\textbf{f})=|\mathcal{A}|
m^{|\mathcal{A}|-1}\Theta(u)\delta\left(\sum_{s\in\mathcal{A}}
f_s-1\right)\left[\prod_{s\not\in\mathcal{A}} \delta_{f_s,0}\right]P[
\textbf{x}^{+}(u,\{f_s\}_{s\in\mathcal{A}})]\nonumber\\
\times\bracket{\prod_{s\not\in
\mathcal{A}}\Theta\left(u+\alpha\epsilon_{s}-y_{s}
\right)}_{\textbf{y}}
\end{eqnarray}
\item \underline{Case $u=0$}. We now must take into account the fact
that $\sum_{s\in\mathcal{A}}f_s+\varphi=1$ with $0\leq \varphi \leq
1$. The stationary equations become
\begin{eqnarray}
x_{s}&=&x^{0}_s\left(\{f_s\}_{s\in\mathcal{A}}\right)\,,\quad\forall
s\in\mathcal{A}\\ y_{s}&<&\alpha\epsilon_{s}\,,\quad\forall s\not
\in\mathcal{A}
\end{eqnarray}
with
\begin{equation}
x^{0}_s\left(\{f_s\}_{s\in\mathcal{A}}\right)\equiv m
f_s+\alpha\epsilon_{s}\,.
\end{equation}
The set of equations $x_s=x_s^{0}(\{f_s\}_{s\in\mathcal{A}})$ defines
an invertible mapping whose Jacobian is
\begin{equation}
\left|\frac{\partial \{x^{0}_s\}_{s\in\mathcal{A}}}{\partial
(\{f_s\}_{s\in\mathcal{A}})}\right|= m^{|\mathcal{A}|}
\end{equation}
Therefore the contribution to the frequency distribution in this cases
reads
\begin{eqnarray}
\hspace{-2.2cm}\varrho_{\mathcal{A}}(u=0,\textbf{f})&=&
m^{|\mathcal{A}|}\delta(u)\int_0^1
d\varphi\,\delta\left(\sum_{s\in\mathcal{A}}f_s+\varphi- 1\right)
\left[\prod_{s\not\in\mathcal{A}}\delta_{f_s,0}\right]
P[\textbf{x}^{0}(\{f_s\}_{s\in\mathcal{A}})]\nonumber\\
\hspace{-2.2cm}&&\times\bracket{\prod_{s\not\in \mathcal{A}}
\Theta\left(\alpha\epsilon_{s}-y_{s}\right)}_{\textbf{y}}\nonumber
\end{eqnarray}
\item \underline{Case $u<0$}. Finally, if all score velocities are
negative then the agent is not on the market and therefore $f_s=0$ for
all $s\in\mathbb{Z}_S$ with
\begin{equation}
u_{s}=\overline{\eta}_{s}-\alpha\epsilon_{s}\,,\quad \forall s\in\mathbb{Z}_S
\end{equation}
and correspondingly
\begin{equation}
\rho(u<0,\textbf{f})\equiv \rho_{{\rm
out}}(\textbf{f})=\prod_{s\in\mathbb{Z}_S} \delta_{f_s,0}
\int_{-\infty}^{0} \frac{du}{\sqrt{2\pi\varsigma^2}}e^{-\frac{(u+
\alpha\epsilon_s)^2}{2\varsigma^2}}
\end{equation}
As was easily expected, the probability that an agent stays out of the
market decreases as $S$ increases, which reflects the simple fact that
the availability of larger strategic alternatives increases the
likelihood that an agent has a profitable strategy among his pool.
\end{enumerate}

Gathering these contributions we finally obtain the probability
distribution of the frequencies and velocity for the subset
$\mathcal{A}$ of strategies of active players and the fraction
$\phi_{{\rm out}}$ of inactive players
\begin{eqnarray}
\hspace{-1.5cm}\varrho_{\mathcal{A}}(u,\textbf{f})&=&|\mathcal{A}|
m^{|\mathcal{A}|-1}\Theta(u)\left[\prod_{s\not\in\mathcal{A}}
\delta_{f_s,0}\right]\delta\left(\sum_{s\in\mathcal{A}}f_s-1
\right)P[\textbf{x}^{+}\left(u,\{f_s\}_{s\in\mathcal{A}}\right)]\nonumber\\
\hspace{-1.5cm}&&\times\bracket{\prod_{s\not\in\mathcal{A}}
\Theta\left(u+\alpha\epsilon_s-y_s\right)}_{\textbf{y}}\nonumber\\
\hspace{-1.5cm}&&+m^{|\mathcal{A}|}\delta(u)\int_0^1
d\varphi\,\delta\left(\sum_{s\in\mathcal{A}}f_s+\varphi-1
\right)\left[\prod_{s\not
\in\mathcal{A}}\delta_{f_s,0}\right]P[\textbf{x}^{0}\left(
\{f_s\}_{s\in\mathcal{A}}\right)]\\
\hspace{-1.5cm}&&\times\bracket{\prod_{s\not\in\mathcal{A}}
\Theta\left(\alpha\epsilon_s-y_s\right)}_{\textbf{y}}\nonumber\\
\hspace{-1.5cm}\phi_{{\rm out}}&=&\prod_{s\in
\mathbb{Z}_S}\int_{-\infty}^{0} 
\frac{du}{\sqrt{2\pi\varsigma^2}}e^{-\frac{(u+\alpha
\epsilon_s)^2}{2\varsigma^2}}
\end{eqnarray}
The frequency distribution is simply given by the sum over all
possible partition of $\mathbb{Z}_S$ (empty set not included). Thus
the average over the initial set of Gaussian variables is converted to
average over the frequency distribution
\begin{equation}
\bracket{(\cdots)}_\star=
\sum_{\mathcal{A}\subseteq\mathbb{Z}_S}\int
du\int d\textbf{f}\,\varrho_{\mathcal{A}}(u,\textbf{f})\,(\cdots)
\end{equation}
 Within this framework, the persistent correlation and susceptibility
 read
\begin{eqnarray}
c=\int d\bepsilon\, w(\bepsilon) \sum_{s=1}^S\bracket{
f^2_{s}}_{\star}\\\chi=\frac{1}{\varsigma^2}\int d\bepsilon\,
w(\bepsilon) \sum_{s=1}^S\bracket{ x_{s}(\textbf{f}) f_s}_{\star}
\end{eqnarray}
where the expression $x_{s}(\textbf{f})$ in the expression for the
susceptibility must be understood as
\begin{equation}
x_{s}(\textbf{f})=\left\{
\begin{array}{cc}
x^+_{s}(\textbf{f})\,,&u>0\\
x^0_{s}(\textbf{f})\,,&u=0
\end{array}\right.
\end{equation}
Interesting information is also provided by the fraction $\phi_{{\rm
in}}(\mathcal{A})$ of active agents using a certain subset
$\mathcal{A}$ of strategies
\begin{equation}
\phi_{{\rm in}}(\mathcal{A})=\int d\bepsilon\,w(\bepsilon) \int du\,
d \textbf{f} \,\varrho_{\mathcal{A}}(u,\textbf{f})
\end{equation}
To quantify our findings we now consider the cases for $S=1$ and $2$
explicitly.

\subsection{$S=1$ (the standard GCMG)}

Here the frequency variable represents the frequency with which the
agent invests. Its distribution becomes
\begin{eqnarray}
\varrho(u,f)&=&\delta_{f,1}\Theta(u)\frac{1}{\sqrt{2\pi
\varsigma^2}}e^{-\frac{(u+m+\alpha
\epsilon)^2}{2\varsigma^2}}+m\delta(u)\frac{1}{\sqrt{2\pi
\varsigma^2}}e^{-\frac{(mf+\alpha\epsilon)^2}{2\varsigma^2}}\nonumber\\
\varrho_{{\rm
out}}(f)&=&\delta_{f,0}\int_{-\infty}^{0}\frac{du}{\sqrt{2\pi
\varsigma^2}}e^{-\frac{(u+\alpha\epsilon)^2}{2\varsigma^2}}
\end{eqnarray}
with $\varsigma^2=\alpha c/(1+\chi)^2$. From here we have the
following expression for the persistent correlation, susceptibility
and fraction of active and inactive agents
\begin{eqnarray}
\hspace{-1.7cm}c&=&\int d\epsilon\, w(\epsilon)\left[\frac{1}{2}{\rm
Erfc}\left(\frac{m+\alpha\epsilon}{\sqrt{2\varsigma^2}}\right)+m
\int_0^{1} \frac{df}{\sqrt{2\pi
\varsigma^2}}e^{-\frac{(mf+\alpha\epsilon)^2}{2\varsigma^2}}
f^{2}\right] \\
\hspace{-1.7cm}\chi&=&\int d\epsilon\,
w(\epsilon)\left[\frac{1}{\sqrt{2\pi \varsigma^2}}
e^{-\frac{(m+\alpha\epsilon)^2}{2\varsigma^2}}+\frac{m}{\varsigma^2}
\int_0^{1} \frac{df}{\sqrt{2\pi \varsigma^2}}
e^{-\frac{(mf+\alpha\epsilon)^2}{2\varsigma^2}}(mf +\alpha\epsilon)
f\right]\\
\hspace{-1.7cm}\phi_{{\rm in}}&=&\int d\epsilon\,
w(\epsilon)\left[\frac{1}{2}{\rm
Erfc}\left(\frac{m+\alpha\epsilon}{\sqrt{2\varsigma^2}}\right)+m
\int_0^{1} \frac{df}{\sqrt{2\pi
\varsigma^2}}e^{-\frac{(mf+\alpha\epsilon)^2}{2\varsigma^2}}\right] \\
\hspace{-1.7cm}\phi_{{\rm out}}&=&\frac{1}{2}\int d\epsilon\,w(\epsilon){\rm
Erfc}\left(-\frac{\alpha\epsilon}{\sqrt{2\varsigma^2}}\right)
\end{eqnarray}
where we have obviously that $\phi_{{\rm in}}+\phi_{{\rm out}}=1$
since the probability $\varrho(u,f)$ is indeed normalised. Taking for
the incentives the distribution
\begin{equation}
w(\epsilon)=m_s\delta(\epsilon-\overline{\epsilon})+ (1-m_s)
\delta(\epsilon+\infty)
\end{equation}
where $m_s$ denotes the fraction of speculators and $1-m_s$ that of
producers, one easily sees that the above equations coincide with
those derived for the GCMG (see {\it e.g.}  \cite{coolen,memory}).

\subsection{$S=2$}

\begin{figure}
\begin{center}
\includegraphics*[height=10cm]{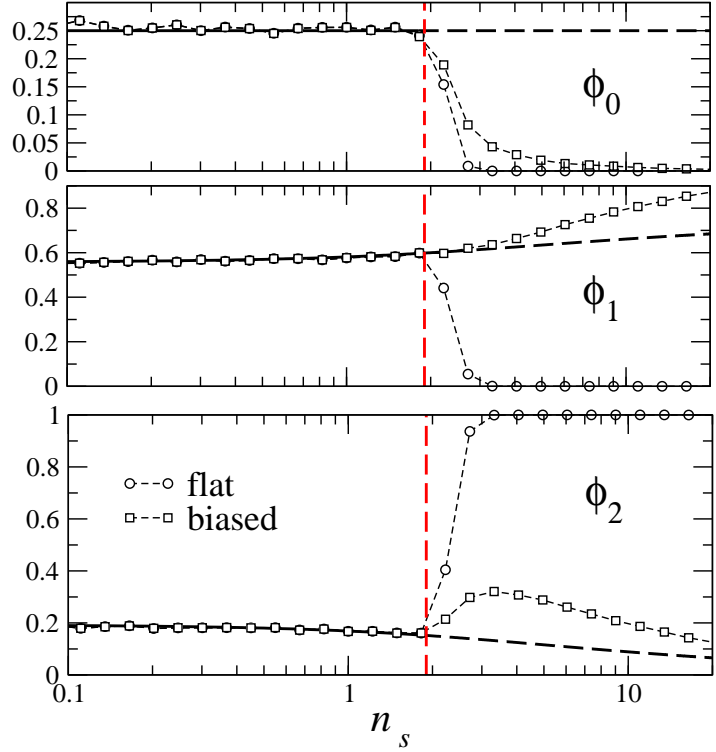}
\caption{\label{gc}Top to bottom: the fraction of speculators using
  $0$, $1$ and both of their strategies versus $n_s$ at $n_p=1$.
  Markers denote results of on-line simulations of systems with
  $N_sP=10^4$ averaged over $200$ disorder samples per point. `Flat'
  refers to initial conditions with $U_{i,1}(0)=U_{i,2}(0)$ for all
  speculators $i$. `Biased' denotes instead initial states with
  $U_{is}(0)=O(\sqrt{N})>0$ and $U_{is'}(0)=0$.  Continuous lines are
  analytic results, and they have been continued as dashed lines in
  the non-ergodic region. The dotted lines joining the markers are a
  guide for the eye. The dashed vertical line marks the critical point
  $n_s\simeq 1.88$ above which the ergodic theory breaks down.}
\end{center}
\end{figure}

The stationary state equations now take the form (\ref{eq:SS}), with
$\mathbb{Z}_S=\{1,2\}$. It is now convenient to consider the following
cases in detail.
\begin{enumerate}
\item \underline{Case $u_s=u\geq 0$ for each $s$.} 
\begin{itemize}
\item If $u>0$, then $f_1+f_2=1$ and
\begin{equation}
u=\overline{\eta}_s-\alpha\epsilon_s-mf_s
\end{equation}
Inverting this mapping we obtain a contribution to the probability
distribution which reads
\begin{equation}
\varrho_1(u,\textbf{f})=2m\Theta(u)\delta\left(f_1+f_2-1\right)
P[\textbf{x}^{+}(u,\{f_s\}_{s\in\mathbb{Z}_2})]
\end{equation}
with $x^+_s(u,f_s)=u+\alpha\epsilon_s+mf_s$.
\item If $u=0$, we have $\varphi+f_1+f_2=1$ with $\varphi\neq 0$ and
correspondingly
\begin{equation}
\fl\varrho_2(u,\textbf{f})=m^2\delta(u)\int_0^1
d\varphi\delta\left(f_1+f_2+\varphi-1\right)
P[\textbf{x}^{0}(\{f_s\}_{s\in\mathbb{Z}_2})]
\end{equation}
with $x^0_s(f_s)=\alpha\epsilon_s+mf_s$.
\end{itemize}
\item \underline{Case $u_s=u>u_s'$ for $s\neq s'$ with $u\geq
0$}. Proceeding as before:
\begin{itemize}
\item If $u>0$, then $f_s=1$ and $f_{s'}=0$, so that
\begin{equation}
\fl\varrho_3(u,\textbf{f})=\Theta(u)\delta_{f_s,1}\delta_{f_{s'},0}
P[x_s^+(u)]\int_{-\infty}^u\frac{1}{\sqrt{2\pi\varsigma^2}}
e^{-\frac{(u_{s'}+\alpha\epsilon_{s'})^2}{2\varsigma^2}}du_{s'}
\end{equation}
with $x_s^+(u)=u+\alpha\epsilon_s+m$.
\item If $u=0$ we have instead
\begin{equation}
\fl\varrho_4(u,\textbf{f})=m\delta(u)\delta_{f_{s'},0}\int_0^1
d\varphi \delta(f_s+\varphi-1)
P[x_s^0(f_s)]\int_{-\infty}^0\frac{1}{\sqrt{2\pi\varsigma^2}}
e^{-\frac{(u_{s'}+\alpha\epsilon_{s'})^2}{2\varsigma^2}}du_{s'}
\end{equation}
with $x^0_s(f_s)=mf_s+\alpha\epsilon_s$.
\end{itemize}
\item \underline{Case $u_1,u_2<0$}. Now $\varphi=1$. This happens with
probability
\begin{equation}
\fl\phi_{{\rm out}}\equiv\phi_0=\int d\textbf{f}
\prod_{s\in\mathbb{Z}_2}\delta_{f_s,0}
\int_{-\infty}^0\frac{1}{\sqrt{2\pi\varsigma^2}}
e^{-\frac{(u_{s}+\alpha\epsilon_{s})^2}{2\varsigma^2}}du_{s}
=\prod_{s\in\mathbb{Z}_2}\frac{1}{2}\left(
1+\erf\frac{\alpha\epsilon_s}{\sqrt{2\varsigma^2}}\right)
\end{equation}
\end{enumerate}
As usual, we divide the population of $N$ agents into two groups,
speculators and producers. As before, the $N_p$ producers have only
one strategy and play at every time step (adopting the notation of
\cite{gcmg}, we write $n_p=N_p/P$), whereas the $N_s$ speculator have
$2$ strategies each (we write $n_s=N_s/P$). The equations for $c$,
$\chi$ and the fraction $\phi_a$ of speculators using $a$ strategies
($a\in\{0,1,2\}$) take a simpler form when, for speculators,
$\epsilon_s=0$ for each $s$. In this case, for the quantity
$y=\sqrt{\alpha/c}$ ($\alpha=P/N$ with $N=N_s+N_p$) and $\chi$ one
finds
\begin{eqnarray}
\fl y^2+n_s \Bigg\{\left(\frac{y^2}{4}+\frac{1}{2}\right)
\erf\frac{y}{2}\erfc\frac{y}{2}
-\frac{y}{2\sqrt{\pi}}\exp\left(-\frac{y^2}{4}\right)
+\frac{3}{4}\left(\erf\frac{y}{2}\right)^2\\\nonumber\label{73}
-\frac{y}{\sqrt{2 \pi}}\exp\left(-\frac{y^2}{2}\right)+
\frac{1}{2}\erf\frac{y}{\sqrt{2}}\Bigg\} =1
\end{eqnarray}
whereas
\begin{eqnarray}
\fl\frac{\chi}{n_s(1+\chi)}=
\frac{1}{2}\erf\frac{y}{2}\erfc\frac{y}{2} -\frac{y}{2\sqrt{\pi}}
\exp\left(-\frac{y^2}{4}\right)\erfc\frac{y}{2}
+\frac{1}{2}\left(\erf\frac{y}{2}\right)^2\nonumber\\
-\frac{y\exp\left(-\frac{y^2}{2}\right)}{\sqrt{2 \pi}}+
\frac{1}{2}\erf\frac{y}{\sqrt{2}}\\
\phi_0=\frac{1}{4}\\ \phi_1=2\int du
d\mathbf{f}\left[\varrho_3(u,\textbf{f})+\varrho_4(u,\textbf{f})\right]
=\frac{1}{2}+J(y)\\ \phi_2=\int du
d\mathbf{f}\left[\varrho_1(u,\textbf{f})+\varrho_2(u,\textbf{f})\right]
=\frac{1}{4}-J(y)
\end{eqnarray}
with
\begin{equation}
J(y)=\frac{1}{\sqrt{2\pi}}\int_0^\infty
\exp\left[-\frac{(x+y)^2}{2}\right]
\erf\frac{x}{\sqrt{2}}dx
\end{equation}
(Note that $J(y)\in[0,1/4]$. Furthermore, $\phi_0+\phi_1+\phi_2=1$.)
Solving (\ref{73}) for $y$ all other quantities can be immediately
evaluated. Fig. \ref{gc} reports the behaviour of $\phi_0$, $\phi_1$
and $\phi_2$ as a function of $n_s$ for $n_p=1$.
The point where simulations and theory depart can be computed assuming
that $\chi\to\infty$ (implying the onset of anomalous response). This
gives the critical point $n_s\simeq 1.88$, above which the ergodicity
assumptions fail and the steady state depends on initial
conditions. Thus this model displays the standard phase transition
with ergodicity breaking characterizing the original $S=1$
GCMG. Similarly to what happens in the canonical MG, the critical
point (which in general depends on $n_p$), decreases as $S$ increases,
a reflection of the fact that agents .

\section{Generalisation to models with $K$ assets}
\label{sect:MG_K_assets}

Multi-asset Minority Games have been introduced in \cite{ma} but we
shall discuss here a slightly more general version of the same
model. One considers a market with $K$ assets
$\sigma\in\mathbb{Z}_K\equiv\{1,\ldots, K\}$ and $N$ agents. At each
time step $\ell$, agents receive $K$ information patterns
$\mu_\sigma(t)\in\{1,\ldots, P_\sigma\}$ chosen randomly and
independently for each $\sigma$ with uniform probability and, based on
these, they formulate their bids (one bid per asset at each time
step). $P_\sigma$ is taken to scale linearly with $N$ and we will
denote their ratios as $\alpha_\sigma=P_\sigma/N$. For each asset
$\sigma$, every agent disposes of $S$ trading strategies
$\{a_{is\sigma}^{\mu_\sigma}\}_{s=1}^S$ that prescribe a binary action
$a^{\mu_\sigma}_{is\sigma}\in\{-1,1\}$, drawn randomly and uniformly,
and independently for each asset, strategy and pattern. The
performance of every strategy foe each asset is monitored by a score
function $U_{is\sigma}(\ell)$ which is updated by the following rule
\begin{equation}\label{online}
U_{is\sigma}(\ell+1)=U_{is\sigma}(\ell)-
a_{is\sigma}^{\mu_\sigma(\ell)}A_{\sigma}(\ell)-\epsilon_{is\sigma}/\sqrt{N}
\end{equation}
where $\epsilon_{is\sigma}$ are real constants representing positive
or negative incentives for the agents to trade, and $A_\sigma(\ell)$
is the excess demand of asset $\sigma$ at time $\ell$,
\begin{equation}
A_\sigma(\ell)=\frac{1}{\sqrt{N}}\sum_{j=1}^Nb_{j\sigma}(\ell)
\end{equation}
where $b_{i\sigma}(\ell)$ denotes the bid formulated by agent $i$ in
asset $\sigma$ at time $\ell$. Let
$\{s_{i\sigma}(\ell)\}_{\sigma\in\mathbb{Z}_K}$ be the strategies he
chooses for each asset and let
$\mathcal{T}_i(\ell)\subseteq\mathbb{Z}_K$ denote the subset of assets
in which agent $i$ trades at time $\ell$. We then write the bid
explicitly in the following form:
\begin{equation}
b_{i\sigma}(\ell)=\sum_{s=1}^S\unity_{\sigma\in\mathcal{T}_i(\ell)}
a_{is\sigma}^{\mu_\sigma(\ell)}\delta_{s,s_{i\sigma}(\ell)}n_{is\sigma}(\ell)
\end{equation}
Here, the terms
$a_{is\sigma}^{\mu_\sigma(\ell)}\delta_{s,s_{i\sigma}(\ell)}n_{is\sigma}(\ell)$
preserve the meaning they had in the single-asset model. The new term
\begin{equation}
\unity_{\sigma\in\mathcal{T}}=\left\{\begin{array}{cl}
1&\sigma\in\mathcal{T}\subseteq \mathbb{Z}_K\\
0&{\rm otherwise}
\end{array}\right.
\end{equation}
defines the set of assets in which agent $i$ is active. We assume now
that 
\begin{eqnarray}
\mathcal{T}_i(\ell)= h_i[\{U_{is\sigma}(\ell)\}]\\
s_{i\sigma}(\ell)= g_i[\{U_{is\sigma}(\ell)\}]
\end{eqnarray}
with $\{g_i\}$ and $\{h_i\}$ generic functions describing the strategy
and asset selection rule. In the model described in \cite{ma}, $S=1$
and $K=2$ with
$\mathcal{T}_i(\ell)=\{\widetilde{\sigma}\in\mathbb{Z}_2{\rm
~s.t.~}\widetilde{\sigma}={\rm arg~}\max_\sigma
U_{is\sigma}(\ell)\delta_{s,s_{i\sigma}(\ell)}\}$.

The batch dynamics can be analysed in terms of $SK$ effective
processes for a single representative agent:
\begin{eqnarray}
\hspace{-1.5cm}U_{s\sigma}(t+1)&=&U_{s\sigma}(t)+
\theta_{\sigma}(t)-\alpha_\sigma\epsilon_{s\sigma}\nonumber\\
&&-\alpha_\sigma\sum_{t'\leq t}
[\unity+G_\sigma]^{-1}(t,t')n_{s\sigma}(t')
\unity_{\sigma\in\mathcal{T}(t')}\delta_{s_\sigma(t'),s}+\eta_{s\sigma}(t)\,,
\end{eqnarray}
where $\{\eta_{s\sigma}(t)\}$ is again a coloured Gaussian noise,
\textit{viz.}
\begin{eqnarray}
\bracket{\eta_{s\sigma}(t)}_\star=0\\
\bracket{\eta_{s\sigma}(t)\eta_{s'\sigma'}(t')}_\star
=\delta_{s,s'}\delta_{\sigma,\sigma'}\alpha_\sigma[(\unity+G_\sigma)^{-1}
C_\sigma (\unity+G^\dag_\sigma)^{-1} ](t,t')
\end{eqnarray}
and where
\begin{eqnarray}
\fl C_{\sigma}(t,t')=\sum_{s=1}^S \int d\bepsilon\,
w(\bepsilon)\int dgdh\,W[g,h]\,
\bracket{n_{s\sigma}(t)n_{s\sigma}(t')\delta_{s,s_\sigma(t)}
\delta_{s,s_\sigma(t')}\unity_{\sigma\in\mathcal{T}(t)}
\unity_{\sigma\in\mathcal{T}(t')}}_\star\label{cgen}\\ \fl
G_{\sigma}(t,t')=\sum_{s=1}^S\int d\bepsilon\, w(\bepsilon) \int
dgdh\,W[g,h]\,
\frac{\delta\bracket{n_{s\sigma}(t)\delta_{s,s_\sigma(t)}
\unity_{\sigma\in\mathcal{T}(t)}}_\star}{\delta
\theta_{s\sigma}(t')}\label{chigen}
\end{eqnarray}
are identified with the bid autocorrelation and response functions of
asset $\sigma$:
\begin{eqnarray}
C_\sigma(t,t')=\frac{1}{N}\sum_{i=1}^N
\left[\bracket{b_{i\sigma}(t)b_{i\sigma}(t')}\right]_{dis}\\
G_\sigma(t,t')=\frac{1}{N}\sum_{i=1}^N
\left[\frac{\partial\bracket{b_{i\sigma}(t)}}{
\partial\theta_{i\sigma}(t')}\right]_{dis}
\end{eqnarray}
in the limit $N\to\infty$. In the above formulas, $W[g,h]$ generalizes
(\ref{densa}) to include the function $h_i$:
\begin{equation}
W[g,h]=\frac{1}{N}\sum_{i\in\mathbb{Z}_N}\delta_{(F)}(g- g_i)
\delta_{(F)}(h- h_i)
\end{equation}

Proceeding as before, one arrives (with obvious notation) at the
following stationary state process:
\begin{equation}
u_{s\sigma}=\overline{\theta}_\sigma+\overline{\eta}_\sigma
-\alpha_\sigma\epsilon_{s\sigma} -m_\sigma
\sum_{\mathcal{T}\subseteq\mathbb{Z}_K}\sum_{n\in\mathcal{I}}
n f_{ns\sigma}(\mathcal{T})
\end{equation}
with
\begin{eqnarray}
m_\sigma=\frac{\alpha_\sigma}{1+\chi_\sigma},~~~~~
\bracket{\overline{\eta}_{s\sigma}^2}\equiv\varsigma_\sigma^2=
\frac{\alpha_\sigma c_\sigma}{(1+\chi_\sigma)^2}\\
f_{ns\sigma}(\mathcal{T})=\lim_{\tau\to\infty}
\frac{1}{\tau}\sum_{t=0}^{\tau-1}\delta_{\mathcal{T},\mathcal{T}(t)}
\unity_{\sigma\in\mathcal{T}}
\delta_{n,n_{s\sigma}(t)}\delta_{s_\sigma(t),s}
\end{eqnarray}
and where the asset-dependent persistent autocorrelation and
susceptibility are given by
\begin{eqnarray}
\fl c_\sigma&=&\sum_{s=1}^S
\sum_{\mathcal{T},\mathcal{T}'\subseteq\mathbb{Z}_K}
\sum_{n,n'\in\mathcal{I}}n\,n'\int d\bepsilon\, w(\bepsilon)
\int dg dh\,W[g,h]\,
\bracket{f_{ns\sigma}(\mathcal{T})f_{n's\sigma}(\mathcal{T}')}_{\star}\\
\fl \chi_{\sigma}&=&\frac{1}{\varsigma^2_\sigma}
\sum_{s=1}^S\sum_{\mathcal{T}\subseteq\mathbb{Z}_K}\sum_{n\in\mathcal{I}}
n\int d\bepsilon\, w(\bepsilon) \int dfdg\,W[f,g]\,
\bracket{\overline{\eta}_{s\sigma} f_{ns\sigma}(\mathcal{T})}_{\star}
\end{eqnarray}
Given a subset of assets $\mathcal{T}$, then
$f_{ns\sigma}(\mathcal{T})$ is the frequency of the asset
$\sigma\in\mathcal{T}$ being traded by using the strategy $s$ when an
action $n$ has been taken on the market. The normalization now reads
\begin{equation}
\sum_{\mathcal{T}\in\mathbb{Z}_K}\frac{1}{|\mathcal{T}|}
\sum_{\sigma\in
\mathcal{T}}\sum_{n\in\mathcal{I}}\sum_{s\in\mathbb{Z}_S}
f_{ns\sigma}(\mathcal{T})=1
\end{equation}

Let us discuss the simplest case in which $F\equiv 1$ and $S=1$,
corresponding to the canonical multi-asset MG (whose particular case
$K=2$ is the subject of \cite{ma}). Agents have at their disposal a
set of $K$ assets to trade, one each time (\textit{i.e.}
$|\mathcal{T}|=1$). We assume that
$\epsilon_{s\sigma}=\epsilon_\sigma$ and that the asset selected at
time $t$ is given by
\begin{equation}
\sigma(t)=h[\{U_\sigma(t)\}]={\rm arg\,
max}_{\sigma}[\{u_{\sigma}(t)\}]
\label{eq:rule}
\end{equation}
Following the same line of arguments one obtains the following
expression for the distribution of frequencies for a subset of assets
$\mathcal{T}$ being traded in the steady state:
\begin{eqnarray}
\fl \varrho_{\mathcal{T}} (u,\textbf{f})=
\sum_{(\sigma_1,\ldots, \sigma_{|\mathcal{T}|-1})
\subset\mathcal{T}}m_{\sigma_1} \cdots m_{\sigma_{|\mathcal{T}|-1}}
\delta\left( \sum_{\sigma\in\mathcal{T}}
f_\sigma-1\right)\left[\prod_{\sigma\not\in\mathcal{T}}
\delta_{f_\sigma,0}\right] \nonumber\\
\times P[\textbf{x}(u,\{
f_\sigma\}_{\sigma\in\mathcal{T}})]\bracket{\prod_{\sigma\not\in
\mathcal{T}}\Theta \left(u-y_{\sigma}\right)}_{\textbf{y}}
\end{eqnarray}
where
\begin{equation}
x_\sigma\left(u,\{f_\sigma\}_{\sigma\in\mathcal{T}}\right)\equiv
u+m_\sigma f_\sigma-\alpha_\sigma\epsilon_\sigma
\end{equation}
As before the frequency distribution is given by the sum over all
possible partitions of $\mathbb{Z}_K$ (empty set not included):
\begin{equation}
\bracket{(\cdots)}_\star=
\sum_{\mathcal{T}\subseteq\mathbb{Z}_K|\mathcal{T}\neq \emptyset}\int
du\, d\textbf{f}\,\varrho_{\mathcal{T}}(u,\textbf{f})\,(\cdots)
\end{equation}
Within this framework, the persistent correlation and susceptibility
read
\begin{eqnarray}
c_\sigma&=&\sum_{\mathcal{T}\subseteq\mathbb{Z}_K|\mathcal{T}\neq
\emptyset}\int du\,d\textbf{f}\,\varrho_{\mathcal{T}}(u,\textbf{f})
f^2_{\sigma}\,,\\ \chi_\sigma&=&\frac{1}{\varsigma^2_\sigma}
\sum_{\mathcal{T}\subseteq\mathbb{Z}_K|\mathcal{T}\neq \emptyset}\int
du\, d\textbf{f}\,\varrho_{\mathcal{T}}(u,\textbf{f})\,
[x_{\sigma}(u,\{f_\sigma\}_{\sigma\in\mathcal{T}}) f_\sigma]
\end{eqnarray}
whereas the fraction $\phi_{\mathcal{T}}$ of agents trading a certain
subset $\mathcal{T}$ reads
\begin{equation}
\phi_{\mathcal{T}}=\int du\, d \textbf{f}
\,\varrho_{\mathcal{T}}(u,\textbf{f})
\end{equation}
It is easily checked that $\phi_{\mathcal{T}}$ satisfies
$\sum_{\mathcal{T}\subseteq\mathbb{Z}_K|\mathcal{T}
\neq\emptyset}\phi_{\mathcal{T}}=1$.

\section{Summary and outlook}

Minority Games with $S$ strategies and/or $K$ assets per agent are
intriguing generalizations of the standard MG setup which display a
qualitatively similar global physical picture (e.g.  regarding the
transition with ergodicity breaking) but substantially richer patterns
of agent behaviour, directly related to the enlargement of the agents'
strategic endowments. The precise characterization of this aspect,
even in the ergodic regime, poses challenging technical problems which
have started to be analysed only recently. The central issue concerns
the calculation of the statistics of the frequencies with which
subsets of strategies are used. This problem was first tackled in
\cite{shayeghi}, where an explicit solution is derived in the context
of canonical batch MGs. In this work we have presented an alternative
and mathematically simpler solution method (though the complexity of
the calculations still increases rapidly with $S$). We have shown
specifically how to recover the theory of the canonical case and
solved explicitly the grand-canonical batch MG with $S$ strategies per
agent. The method also generalizes to the recently introduced
multi-asset models.

The method discussed here can be applied to a number of variants of
the basic setup, some of which may be important from an economic
viewpoint (for example in order to study the emergence of cross-asset
correlations). Its main limitation is that, while the effective-agent
dynamics, eqn. (10), holds true in both the ergodic and non-ergodic
phases, our futher focus on time-translational properties limits the
rigour of our conclusions to the ergodic regime. The richness of the
MG dynamics is actually most striking when ergodicity is broken.
Multi-strategy MGs are likely to produce a variety of possible steady
states that may require novel observables to be completely
characterised. Up to now, our understanding of non-ergodic regimes
relies entirely on {\it ad hoc} heuristic arguments (see for instance
\cite{batch,ctl}) which provide a rough picture of the geometry of
steady states and of the role of initial conditions for obtaining
states of high or low volatility, but a more precise characterization
remains elusive. In our opinion, at the present stage of our
theoretical understanding of MGs, any advance in this direction would
be most welcome.

~

\ack It is a pleasure to thank ACC Coolen and N Shayeghi for useful
discussions. We acknowledge financial support from the EU grant
HPRN-CT-2002-00319 (STIPCO).

~

~


\section*{References}

\begin{thebibliography}{99}

\bibitem{mgbook}Challet D, Marsili M and Zhang YC 2005 Minority Games
(Oxford University Press, Oxford)

\bibitem{coolen}Coolen ACC 2005 The mathematical theory of Minority
Games (Oxford University Press, Oxford)

\bibitem{shayeghi}Shayeghi N and Coolen ACC 2006 {\it J. Phys. A:
    Math. Gen.} {\bf 39} 13921

\bibitem{gcmg}Challet D and Marsili M 2003 {\it Phys. Rev. E} {\bf 68}
036132

\bibitem{ma}Bianconi G, De Martino A, Ferreira FF and Marsili M 2006
Preprint physics/0603152

\bibitem{reviu}De Martino A and Marsili M 2006 {\it J. Phys. A: Math.
    Gen.} {\bf 39} R465

\bibitem{noiz}Coolen ACC, Heimel JAF and Sherrington D 2001 {\it
Phys. Rev. E} {\bf 65} 016126

\bibitem{batch}Heimel JAF and Coolen ACC 2001 {\it Phys. Rev. E} {\bf
63} 056121

\bibitem{dedo}De Dominicis C 1978 {\it Phys. Rev. B} {\bf 18} 4913

\bibitem{memory}Challet D, De Martino A, Marsili M and Perez Castillo
I 2006 JSTAT P03004

\bibitem{ctl}Marsili M and Challet D 2001 {\it Phys. Rev. E} {\bf 64}
  056138

\end{thebibliography}
\end{document}